\documentclass{ws-procs9x6}

\def\lsim{\raise0.3ex\hbox{$<$\kern-0.75em\raise-1.1ex\hbox{$\sim$}}}
\def\gsim{\raise0.3ex\hbox{$>$\kern-0.75em\raise-1.1ex\hbox{$\sim$}}}

\begin{document}

\title{Renormalized quark-anti-quark free energy
\footnote{\uppercase{S}upported in part by
    the \uppercase{DFG} under grant \uppercase{FOR} 339/1-2. }}

\author{F. Zantow$^1$ \lowercase{with} O. Kaczmarek$^1$, F. Karsch$^1$,
  P. Petreczky$^2$\\}

\address{$^1$Fakult\"at f\"ur Physik, Universit\"at Bielefeld, D-33615 Bielefeld,
  Germany\\
E-mail: zantow@physik.uni-bielefeld.de}
\address{$^2$Brookhaven National Laboratory, Physics Department, Bldg. 510\\
  Upton, NY 11973-5000}

\maketitle

\abstracts{We present results on the renormalized quark-anti-quark free energy
  in $SU(3)$ gauge theory at finite temperatures. We discuss results for the
  singlet, octet and colour averaged free energies and comment on thermal
  relations which allow to extract separately the potential energy and entropy
  from the free energy. }
\section{Introduction}
It is believed that strongly interacting matter undergoes a
confinement-deconfinement phase
transition at very high temperature $T$ and/or baryon density and it is
well-established that QCD is a valid theory for both phases. The thermal
modification of
meson and hadron properties is often discussed in terms of $n$-point
Polyakov loop
correlation functions which describe the free energy $F_{nq,{\tilde
    n}\bar{q}}$ of a gluonic
heat bath including static colour charges\cite{McLerran}. It is well-known,
however, that Polyakov loop
correlation functions need to be fixed by renormalization. On a lattice a
renormalization prescription has recently been
suggested\cite{Kacze}. Applying
this renormalization scheme to the $n$-point Polyakov loop correlation
functions, one fixes the free energy which consists of
potential energy ($V_{nq,{\tilde
    n}\bar{q}}$) and entropy ($S_{nq,{\tilde
    n}\bar{q}}$): $F_{nq,{\tilde
    n}\bar{q}}=V_{nq,{\tilde
    n}\bar{q}}-TS_{nq,{\tilde
    n}\bar{q}}$. The calculation of the potential energy from Polyakov loop
correlation functions at finite $T$ is still an open problem - although
the potential energy is expected to be an essential quantity for heavy-quarkonium
physics.

In this paper we refer to results on the renormalization
scheme\cite{Kacze,Kacze2} applied to
2-point Polyakov loop correlation functions in $SU(3)$ gauge theory on a
lattice. We discuss the renormalized colour singlet, colour octet and colour
averaged quark-anti-quark free energy at temperatures below and above the
critical temperature. We then comment on the properties of the
ingredients of the quark-anti-quark free energy, the potential energy and the
entropy at finite $T$. Further details on the lattice calculations are
presented in Ref.~2.

\section{A new look at quark-anti-quark free energy}
\begin{figure}[t]
\centerline{\epsfxsize=9cm\epsfysize=9cm\epsfbox{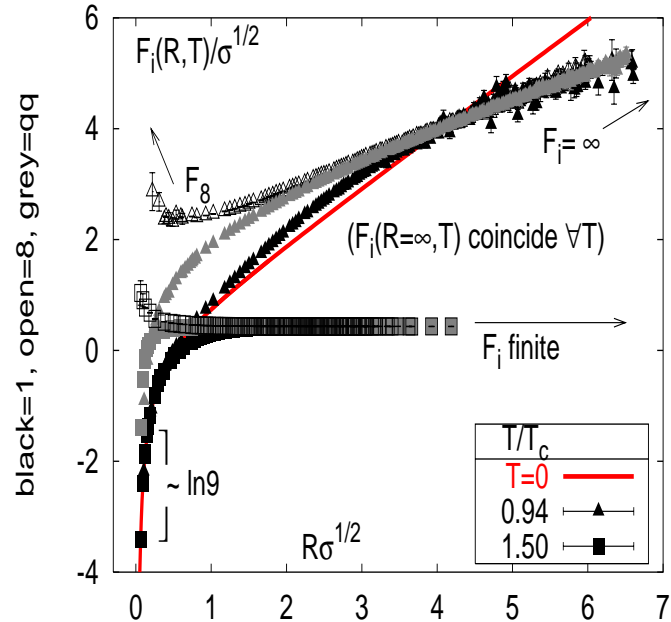}}
\caption{The renormalized quark-anti-quark free energy in $SU(3)$ as a
  function of $R\sigma^{1/2}$
  ($\sigma$ being the string tension at $T=0$). The filled black symbols
  indicate the colour singlet free energy and the open symbols the colour octet
  one. The colour averaged free energy is given by the grey (filled)
  symbols. Free energies at equal temperatures are presented with the same
  symbol-style (triangles and squares). The grey curve is the $T=0$
  potential; At short distances it is given perturbatively$^5$ while it
  coincides at large quark anti-quark separations with the standard Cornell
  potential (see also Ref. 6).
In this figure we have included data from lattice calculations
  on lattices of size $32^3\times4,8,16$ and $64^3\times4$.}
\label{free_energy}
\end{figure}

The renormalized
quark-anti-quark free energies in each colour channel are shown in
Fig.~\ref{free_energy} exemplarily
for one temperature below and one above $T_c$.
Also included is the heavy quark potential at zero
temperature which we take to be the standard Cornell potential. In a rather
rough way we may
distinguish three different distance regions with different behaviour of the
free energies at fixed temperature:

At {\em large quark anti-quark separations}
the colour singlet ($F_1$), colour octet ($F_8$) and colour averaged
($F_{q\bar{q}}$) quark-anti-quark free
energies coincide at both temperatures. The same feature can be observed at
all temperatures analyzed by us. In fact, if the separation between the quark
anti-quark sources becomes large, the relative orientation of the colour charges
in the colour space will not influence the free energy: For large distances we
therefore expect $F_1(R\gg1,T)=F_8(R\gg1,T)=F_{q\bar{q}}(R\gg1,T)$ in both
phases. For $T<T_c$, this picture
suggests that the quark-anti-quark free energy is
dominated by a single, temperature dependent string tension for all color
channels at large
separations - including the color
octet channel.

At {\em intermediate distances}, i.e. for distances $1\lsim
R\sqrt{\sigma}\lsim4$ and
temperatures below $T_c$, we note that all color channels lead to an
enhancement of the free energies compared to
the potential at zero temperature.
In this intermediate distance regime the different colour structures of the free
energies become visible. Both features, the confining colour octet
free energy below $T_c$ and the enhancement of the colour singlet free energy
were first observed\cite{Philipsen} in $SU(2)$ - and recent lattice
studies\cite{Fortunato} of
Polyakov loop correlation functions in $SU(2)$ indicate similar properties.

At {\em small quark anti-quark separations} the different
colour structures of the
free energies dominate the picture: The color
singlet free energy coincides with
the potential at zero temperature while the colour octet free energy behaves
repulsive as one may expect from leading order perturbation theory. The color
averaged free
energy respects the
relation $F_{q\bar{q}}-F_1=T\ln9$ at short distances. It is worth noting that
the temperature dependence of the free energy becomes less important with
decreasing quark anti-quark separations in all colour channels.

\section{From quark-anti-quark free energies to the QCD-force at finite $T$?}
The investigation of medium effects on free energies is
essential in heavy-quark physics and has been subject of many
lattice studies so far. Of even greater relevance for heavy-quark physics then
the free energy, however, is the potential energy at finite temperature as it
appears in fundamental
quantum mechanical and field theoretical relations. Thus the knowledge of
potential energy would lead to a better understanding of the basic
 forces in particle physics - but its calculation on a lattice is still an
 outstanding problem at finite temperatures.

\begin{figure}[t]
\centerline{\epsfxsize=8cm\epsfysize=7cm{\epsfbox{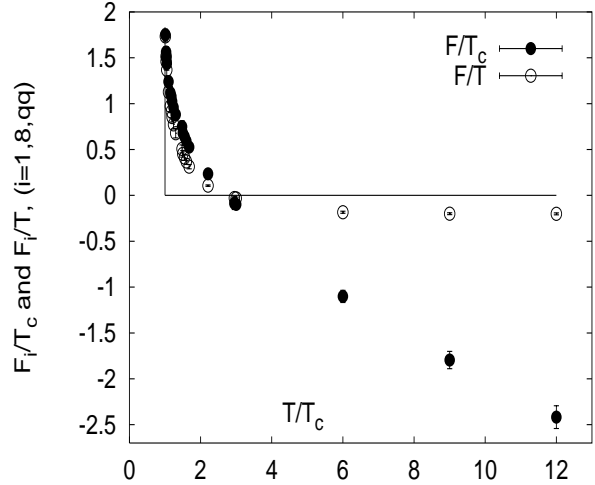}}}
\caption{The renormalized free energy at infinite quark separation for
  $T>T_c$. The filled symbols describe $F^\infty/T_c$ and the open symbols
  $F^\infty/T$. According to our discussion in section 2 the data points reflect the
  free energy for all colour channels. In combination with
  (\ref{relations}) the magnitude of the potential energy can be read off from
  the open symbols while the magnitude of the entropy follows from the filled
  symbols. The same study can be done at any finite quark anti-quark separation
  and will lead to the force at finite temperature. }
\label{potential}
\end{figure}

Having fixed the free energy, however, opens the opportunity\cite{Kacze,Kacze2}
to refer to thermal relations
in order to study, for instance, the potential energy and entropy,
at arbitrary distance $R_0$
\begin{eqnarray}
S(R_0,T)\Big|_{R_0}=-\frac{\partial F}{\partial T}\Big|_{R_0}\qquad\mbox{and }
V(R_0,T)\Big|_{R_0}=-T^2\frac{\partial F/T}{\partial T}\Big|_{R_0};
\label{relations}
\end{eqnarray}
When analyzing these relations at a set of distances, the QCD force at finite
$T$ follows from $-dV/dR$. In order to analyze the quantities in
(\ref{relations}) we
have plotted $F^\infty=\lim_{R\to\infty}F_1(R,T)$ for temperatures above $T_c$ (see
Fig.~\ref{potential}).
We only note here two interesting features that follow from this figure:
Firstly, since $F^\infty/T_c$ is a monotonicly decreasing function with
increasing temperature it
follows that at large quark anti-quark separation the entropy
contribution to the free energy is non-zero (and positive). On the other hand
the free
energies become temperature independent in the limit of
vanishing quark anti-quark separation. This implies that the entropy
contribution to the free energy is $R$-dependent,
\begin{eqnarray}
F_i(R,T)&=&V_i(R,T)-TS_i(R,T) \qquad\mbox{with }i=1,8,q\bar{q}.
\end{eqnarray}
Moreover, the behaviour of $F^\infty/T$ in Fig.~\ref{potential} tells us
that also
the potential energy is non-zero at large quark anti-quark
separations and temperatures near $T_c$. Indeed, in the temperature regime
analyzed by us the potential energy at large quark anti-quark separations
is a positive and monotonicly decreasing function with increasing
temperature. Moreover, from high temperature perturbation theory, one
expects
$F^\infty(T)\sim-Tg^2(T)$ which implies
$V^\infty(T)\sim-2\beta_0Tg^4(T)$ with $\beta_0$
being the first coefficient of the perturbative $\beta$-function.

It should be obvious that the outlined program -
including the renormalization prescription - is neither bound to a temperature
nor to a distance regime and equally well should hold for free
energies in full QCD. It will be of even greater importance in this case as the
potential energy reflects the string breaking energy.

\section{Conclusion}
We have presented a new look at the renormalized quark-anti-quark free
energy
including the colour singlet, colour octet and colour averaged free energy in the
confined and deconfined phase. The renormalization scheme also allows to have a
new interpretation of the heavy quark potential at finite temperatures. In particular, we
have shown that the
$R$-dependence of the potential energy is not only given by that of the free
energy. A
detailed analysis of the potential energy, the entropy and the QCD-force at
finite temperature will be presented elsewhere.

\end{document}